\documentclass[prl,twocolumn,superscriptaddress,nobibnotes,letterpaper]{revtex4-1}

\usepackage[subpreambles=true]{standalone}
\usepackage{afterpage}
\usepackage{graphicx}
\usepackage{dcolumn}
\usepackage{bm}
\usepackage[plain]{fancyref}
\usepackage{placeins}
\usepackage{hyperref}
\usepackage{lipsum}
\usepackage{color,soul}
\usepackage{siunitx}
\usepackage{natbib}
\usepackage{chemformula}
\usepackage{float}

\newcommand{\Ej}{E_\mathrm{J}}
\newcommand{\Ec}{E_{C}}

\newcommand{\fres}{f_\mathrm{res}}
\newcommand{\fbare}{f_\mathrm{bare}}
\newcommand{\tm}{t_\mathrm{-}}
\newcommand{\tp}{t_\mathrm{+}}
\newcommand{\Tmin}{T_\mathrm{-}}
\newcommand{\Tplus}{T_\mathrm{+}}
\newcommand{\Bpar}{B_{\parallel}}
\newcommand{\Bperp}{B_{\perp}}
\newcommand{\gc}{g_\mathrm{c}}
\newcommand{\Tp}{\overline{T}_{P}}
\newcommand{\xqp}{x^0_{\mathrm{QP}}}
\newcommand{\ngg}{n_\mathrm{g}} 
\newcommand{\Tqp}{T_{P}}
\newcommand{\Vc}{V_\mathrm{j}}
\newcommand{\Visl}{V_\mathrm{isl}}
\newcommand{\Vgnd}{V_\mathrm{gnd}}
\newcommand{\Vg}{V_\mathrm{g}}
\newcommand{\fd}{f_\mathrm{d}}
\newcommand{\fp}{f_\mathrm{p}}
\newcommand{\fq}{f_\mathrm{q}}
\newcommand{\fresm}{f_\mathrm{res-}}
\newcommand{\fresp}{f_\mathrm{res+}}
\newcommand{\fpm}{f_\mathrm{p-}}
\newcommand{\fpp}{f_\mathrm{p+}}
\newcommand{\fqm}{f_\mathrm{q-}}
\newcommand{\fqp}{f_\mathrm{q+}}
\newcommand{\EQP}{E_\mathrm{QP}}

\begin{document}
\title{Quasiparticle trapping by orbital effect in a hybrid superconducting-semiconducting circuit}
\author{W. Uilhoorn}
\affiliation{QuTech and Kavli Institute of Nanoscience, Delft University of Technology, 2600 GA Delft, The Netherlands}
\author{J. G. Kroll}
\affiliation{QuTech and Kavli Institute of Nanoscience, Delft University of Technology, 2600 GA Delft, The Netherlands}
\author{A. Bargerbos}
\author{S. D. Nabi}
\affiliation{QuTech and Kavli Institute of Nanoscience, Delft University of Technology, 2600 GA Delft, The Netherlands}
\author{C. K. Yang}
\affiliation{Microsoft Quantum Lab Delft, 2600 GA Delft, The Netherlands}

\author{P. Krogstrup}
\affiliation{Center for Quantum Devices and Microsoft Quantum Materials Lab, Niels Bohr Institute, University of Copenhagen, Copenhagen, Denmark}
\author{L. P. Kouwenhoven}
\affiliation{QuTech and Kavli Institute of Nanoscience, Delft University of Technology, 2600 GA Delft, The Netherlands}
\affiliation{Microsoft Quantum Lab Delft, 2600 GA Delft, The Netherlands}
\author{A. Kou}
\affiliation{Department of Physics and Frederick Seitz Materials Research Laboratory, University of Illinois Urbana-Champaign, Urbana, IL 61801, USA}
\author{G. de Lange}
\affiliation{Microsoft Quantum Lab Delft, 2600 GA Delft, The Netherlands}

\date{\today}

\begin{abstract}
The tunneling of quasiparticles (QPs) across Josephson junctions (JJs) detrimentally affects the coherence of superconducting and charge-parity qubits, and is shown to occur more frequently in magnetic fields.
Here we demonstrate the parity lifetime to survive in excess of \SI{50}{\micro\second} in magnetic fields up to \SI{1}{\tesla}, utilising a semiconducting nanowire transmon to detect QP tunneling in real time.
We exploit gate-tunable QP filters and find magnetic-field-enhanced parity lifetimes, consistent with increased QP trapping by the ungated nanowire due to orbital effects.
Our findings highlight the importance of QP trap engineering for building magnetic-field compatible hybrid superconducting circuits.
\end{abstract}

\maketitle
There is wide interest in understanding the dynamics of quasiparticle (QP) excitations in superconducting devices such as those used in astronomy~\cite{Day2003}, metrology~\cite{Pekola2008}, thermometry and refrigeration~\cite{Giazotto2006}, superconducting quantum devices~\cite{Aumentado2004,Visser2011,vanWoerkom2015}, superconducting qubits~\cite{Sun2012,Riste2013,Wang2014,Janvier2015,Riwar2016,Hays2017,Serniak2019}, and proposed topological qubits~\cite{Karzig2020}.
Despite operation at low temperatures where thermal QPs are exponentially suppressed~\cite{Kaplan1976}, non-equilibrium QPs continue to plague these applications.
In particular for superconducting quantum circuits operated as qubits, the tunneling of QPs results in the loss of quantum information, either due to energy exchange with the circuit~\cite{Sun2012,Riste2013,Wang2014,Riwar2016}, or through the complete poisoning of the charge-parity defined qubit subspace~\cite{Karzig2020}.
To prevent such quasiparticle poisoning (QPP), QPs can be captured away from sensitive regions of the device by engineering QP traps, which generally rely on a local reduction of the superconducting gap~\cite{Sun2012,Wang2014,Riwar2016,Taupin2016}.
Hybrid superconductor-semiconductor nanowire (NW) devices provide a novel platform to apply such gap engineering in-situ~\cite{Menard2019}, enabled by their electric-field-tunable proximity-induced energy gap~\cite{de_Moor_2018}.
Moreover, these hybrid NWs show induced superconductivity at magnetic fields strong enough where they are expected to host Majorana zero modes~\cite{Lutchyn2018}, the basis for topological qubits.
In this context, a crucial outstanding question is - how does QPP develop as magnetic field increases?

Recent studies focusing on QPP in superconducting qubits have been based on real-time detection of QPP by detecting parity switching events, utilizing offset-charge sensitive (OCS) transmon qubits~\cite{Riste2013,Serniak2019} and Andreev levels~\cite{Janvier2015, Hays2017}, and all performed while being heavily shielded from stray magnetic fields.
To date, magnetic field studies of QP dynamics in Josephson junction (JJ) circuits consist only of time-averaged experiments performed below $B<\SI{.3}{\tesla}$. 
These works demonstrated the increase in QPP due to field induced Cooper pair breaking~\cite{vanWoerkom2015,vanVeen2018}.
Recent developments in magnetic field compatible superconducting resonators~\cite{Kroll2019} and advances in semiconducting NW-based JJs enabled novel semiconductor-based superconducting circuits~\cite{DeLange2015, Larssen2015, Luthi2018}, that operate in magnetic fields up to \SI{1}{\tesla}~\cite{PitaVidal2020,Kringhoj2021}.
Such devices enable the investigation of QPP in the unexplored magnetic field parameter space that is essential in topological quantum computing, where a magnetic field as strong as \SI{1}{\tesla} may be required~\cite{Lutchyn2018}.

\begin{figure}
    \centering
    \includegraphics{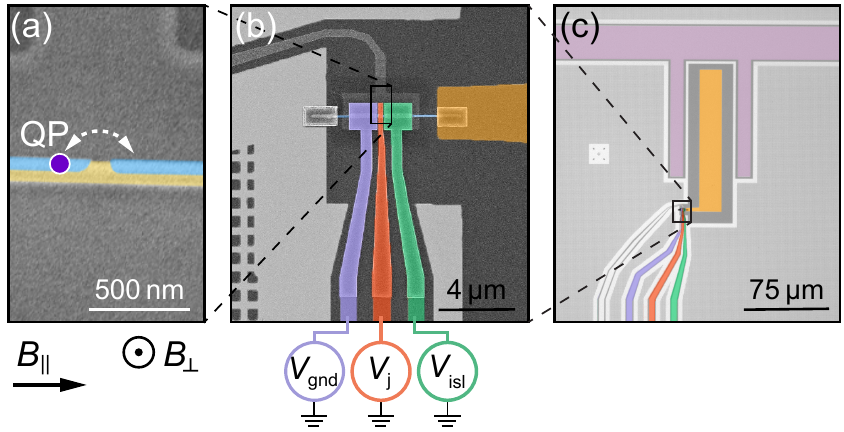}
    \caption{Device overview. (a) False-colored scanning electron micrograph of a lithographically similar device before top gate fabrication, showing the JJ created from a proximitized InAs NW (yellow) with a two-facet epitaxial Al shell (blue). The purple dot illustrates the tunneling of QP excitations across the JJ. (b) False-colored scanning electron micrograph of the NW contacted to ground at the left and to the capacitor island (orange) at the right. A gate (red) at the junction provides $\Ej\left(\Vc\right)$ tuning of the qubits frequency. Plunger gates (green, purple) at either side of the JJ control the chemical potential of the proximitized sections of the NW ($\Visl, \Vgnd$), while also providing a tuning knob for the offset charge $\ngg\left(\Visl\right)$. (c) False-colored optical micrograph of the transmon capacitively coupled to a superconducting transmission coplanar waveguide resonator (pink).}
    \label{fig:device}
\end{figure}

In this letter, we study in-situ tunable gap engineering, and its effect on QPP at magnetic fields up to \SI{1}{\tesla} in an InAs NW JJ using an OCS transmon architecture.
We utilize local gates to electrically control the superconductor-semiconductor coupling and form gate-tunable QP filters situated on either side of the JJ.
When the JJ is protected by the gate-controlled QP filters, we find peaks of enhanced charge-parity lifetimes induced by the magnetic field and centered around odd multiples of $B_\parallel\sim\SI{280}{\milli\tesla}$.
We attribute these enhancements to the formation of quasiparticle traps in the ungated NW segments as a result of a local suppression of the induced gap by the orbital effect~\cite{Buttiker1986, Zuo2017, Winkler2019, Kringhoj2021, Stampfer2021}, efficiently trapping QPs away from the JJ.
Finally we conclude that such devices can be operated at high magnetic field without being limited by QPP, paving the way for implementing proposals for novel superconducting-topological hybrid qubits~\cite{Pekker2013,Ginossar2014}.

Our device is a single-island OCS transmon dispersively coupled to a half-wave resonator, used to read out the state of the qubit~\cite{Koch2007}~\footnote{We measure the exact same device as~\cite{Bargerbos2020}}. The transmon is based on a superconducting-semiconducting (SC-SM) Al/InAs NW JJ [\Fref{fig:device}(a)]~\cite{Krogstrup2015}.
The $\sim\SI{100}{\nano\meter}$ JJ divides the NW in two Al-proximitized InAs sections.
One section is connected to a NbTiN ground plane and the other to a NbTiN island, with charging energy $\Ec \sim h \times \SI{530}{\mega\hertz}$ [\Fref{fig:device}(b)].
The island is capacitively coupled to a superconducting coplanar waveguide resonator for readout with $\fbare = \SI{5.201}{\giga\hertz}$ and $\kappa/2\pi=\SI{2.591}{\mega\hertz}$ with a coupling strength of $g\sim h \times\SI{81}{\mega\hertz}$ [\Fref{fig:device}(c)].
The resonator, ground plane and island are patterned with vortex pinning sites~\cite{Kroll2019}.
The carrier density in the semiconducting JJ and adjacent leads are tuned by several electrostatic gates through the field effect [\Fref{fig:device}(d)]. 
First, we alter the Josephson coupling $\Ej$ by $\Vc$~\cite{Doh2005}, similarly to previous work in the context of SM weak link qubits~\cite{Larssen2015, DeLange2015, Luthi2018}.
In addition, the carrier density profile in the SC-SM parts of the leads are tuned by the wrapped plunger gates $\Visl$ and $\Vgnd$, enabling us to control the proximity effect in the enclosed leads~\cite{Antipov2018,de_Moor_2018,Winkler2019}.
Lastly, the offset charge $\ngg$ of the island is predominantly tuned by $\Visl$.
We note that tuning $\ngg$ requires much smaller ranges in voltage (on the order of $\sim~\SI{100}{\micro\volt}$) than the voltages required to tune the carrier density appreciably (typically requiring $\sim~\SI{100}{\milli\volt}$), providing an effectively independent tuning knob for $\ngg$.
 
\begin{figure}
    \centering
    \includegraphics{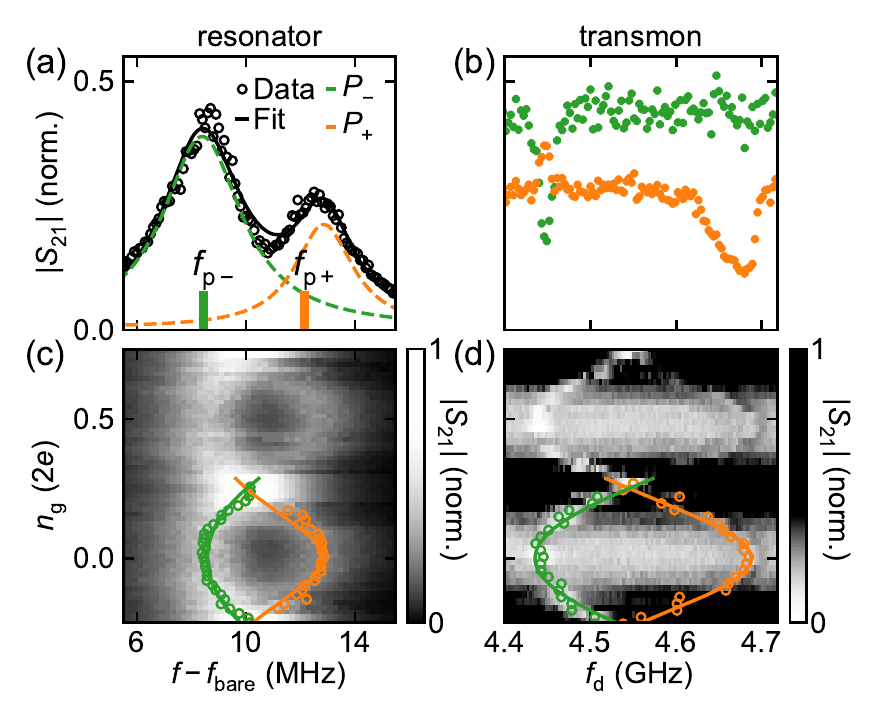}
    \caption{Offset-charge dependent resonator-qubit response, normalised by the maximum transmission at charge degeneracy.  (a) $|S_{21}|$ of the resonator coupled to an $\ngg$ dependent transmon at maximum charge dispersion. The bimodal spectrum is fitted with a double Lorentzian (solid black), and decomposed in the two parity contributions (dashed green, orange). (b) A qubit state dependent resonator shift is observed when driving the qubit by a second input tone $\fd$ while probing the resonator response at a fixed readout frequency. At fixed readout frequency $f_\mathrm{p-}$ ($f_\mathrm{p+}$), a dip corresponding to the $P_-$ ($P_+$) branch of the transmon is observed, with $P_-$ ($P_+$) indicated in green (orange). (c) $|S_{21}|$ versus $\ngg$, revealing the charge modulation directly in the resonator transmission and (d) in qubit spectroscopy probed at $\fp = \fbare +\SI{10}{\mega\hertz}$. The extracted peaks (valleys) are fitted using a standard SIS transmon model resulting in $E_\mathrm{J} = \SI{7.214}{\giga\hertz}$, $\Ec = \SI{0.532}{\giga\hertz}$, $\gc/2\pi = \SI{81}{\mega\hertz}$.}
    \label{fig:resonator_response}
\end{figure}

We first demonstrate readout of the charge-parity $P$ of the OCS transmon via the superconducting microwave resonator response in absence of an additional qubit drive tone~\cite{Serniak2019}.
The $\Ej/\Ec$-ratio is designed to be $\Ej/\Ec\sim 14$ when the qubit's ground to first excited state transition frequency  $f_\mathrm{q}\sim\fbare$, making the qubit transitions depend strongly on parity and $\ngg$ [\Fref{fig:resonator_response}(b,d)]~\cite{Koch2007}.
As a consequence of the strong dispersive coupling between the OCS transmon and resonator, the resonator frequency $\fres$ inherits these same dependencies on $P$ and $\ngg$ [\Fref{fig:resonator_response}].
We verify this by investigating the correspondence between the $P$-dependent resonator frequencies, $\fresm$ and $\fresp$~\footnote{The different magnitudes of $S_{21}$ for $\fresm$ and $\fresp$ are only found with the TWPA activated.}, and the two qubit parity branches, $\fqm$ and $\fqp$, respectively.
By matching the probe frequency $\fpm$ ($\fpp$) to $\fresm$ ($\fresp$) we demonstrate selective sensitivity to $\fqm$ ($\fqp$) in two-tone spectroscopy at maximum charge-parity separation ($\ngg=0$) [\Fref{fig:resonator_response}~(a,b)].
The even and odd charge-parity branches are observed simultaneously, consistent with the transfer of one or more QPs across the JJ during each $\SI{40}{\milli\second}$ measurement.
Our cQED setup allows us to detect individual parity switching events in real time.
The QPP dynamics, and its dependence on electric and magnetic field, is central to this work.

\begin{figure}
    \centering
    \includegraphics{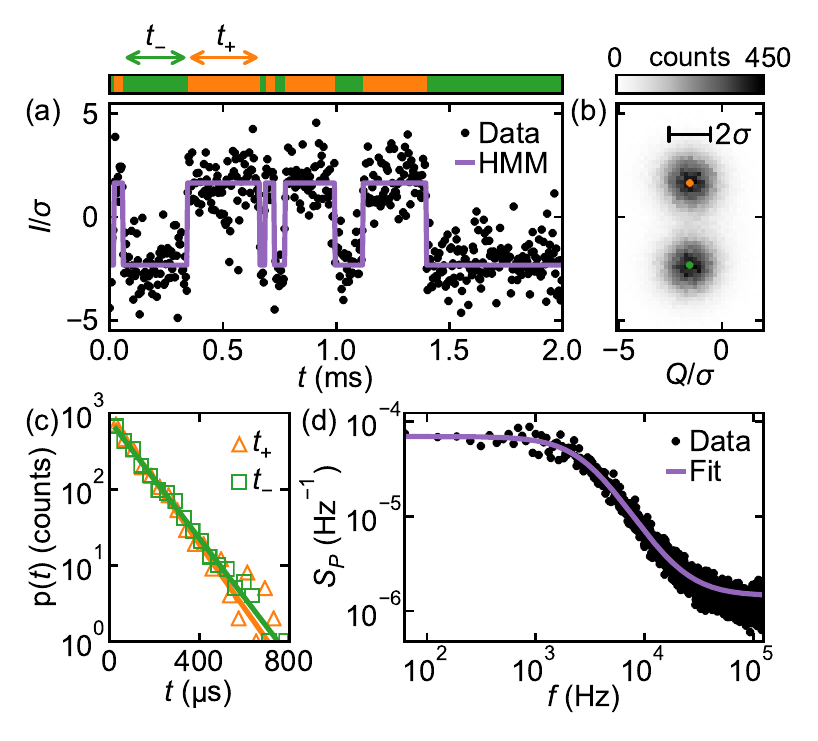}
    \caption{Parity lifetime analysis of $32\times\SI{16}{\milli\second}$ jump traces with $t_\mathrm{int} = \SI{4}{\micro\second}$. (a) \SI{2}{\milli\second} snapshot of a single quadrature response in time (black) revealing a charge parity jump trace. Charge parity state assignment and the extraction of dwell times $\tp$ and $\tm$ (upper panel) is done by applying a hidden Markov model (purple). (b) 2D histogram of the resonator response demonstrating the clustering into two well-separated Gaussian distributions in the IQ-plane; the charge parity states. (c) Histogram of extracted dwell times $\tp$ and $\tm$, corresponding to coherence times of $\Tplus =\SI{104(3)}{\micro\second}$ and $\Tmin =\SI{108(4)}{\micro\second}$ respectively. (d) Average power spectrum density of the 32 independent complex jump traces (black) fitted by a Lorentzian lineshape (purple) corresponding to $\Tqp =\SI{106.9(7)}{\micro\second}$ and a detection fidelity of $F = \SI{0.799(2)}{}$.}
    \label{fig:QPP_Analysis}
\end{figure}

We demonstrate our ability to measure QPP in real time.
By monitoring $S_{21}$ at a fixed readout frequency, we can directly detect island-parity switching events down to $\sim\SI{1}{\micro\second}$ resolution by utilizing a near-quantum limited travelling-wave parametric amplifier~\cite{Macklin2015}. 
To maximize signal to noise, the readout is performed at $\ngg=0$ with $\fp$ set at the local minimum in between the two parity resonances. 
Recording a continuous measurement of $S_{21}$ reveals a random telegraph signal [\Fref{fig:QPP_Analysis}(a)] as a consequence of the switches in qubit-island parity.
The histogram of $S_{21}$ in the complex ($I,Q$)-plane confirms two well-separated parity states~[\Fref{fig:QPP_Analysis}(b)], both Gaussian distributed with symmetric widths and populations~\cite{SI}.
The histogram is built by constructing a connected time series of ($I,Q$)-shots of $S_{21}$ by binning $32\times\SI{16}{\milli\second}$-long time traces in $t_\mathrm{int}=$~\SI{4}{\micro\second} bins.
The equal parity population, as expected with the ground state charge dispersion being two orders smaller than the thermal energy, makes the two parity levels indistinguishable and hinders `even' and `odd' state labelling.
Instead, we refer to `+' and `-' as unspecified parity labels.

To gain further insight in the poisoning dynamics, we assign a state to every ($I,Q$)-shot by applying a hidden Markov model (HMM) to the time evolution of $(I,Q)$ [\Fref{fig:QPP_Analysis}(a)].
The extracted dwell times $\tp$ and $\tm$ are found to follow an exponential probability distribution with average dwell times of $\Tplus =\SI{104(3)}{\micro\second}$ and $\Tmin =\SI{108(4)}{\micro\second}$ [\Fref{fig:QPP_Analysis}(c)].
The exponential distribution indicates that the switching events result from a Poissonian process.
The Poissonian temporal distribution together with the Gaussian distributed parity clusters validate the use of the HMM~\cite{SI}.
The overlap of the dwell time distributions, both in number of counts and average dwell times, demonstrates the symmetry between the even-to-odd and odd-to-even poisoning processes.
This justifies the extraction of a single parity lifetime via Fourier transform analysis [\Fref{fig:QPP_Analysis}(d)]; each \SI{16}{\milli\second} long jump trace is transformed to frequency space separately and subsequently averaged, revealing a power spectral density (PSD) that is well fitted by a Lorentzian~\cite{SI}.
The typical drop off frequency corresponds to the combined average parity lifetime $\Tqp$, where $\Tqp =~\SI{106.9(7)}{\micro\second}$ and a parity detection fidelity of $F = \SI{0.799(2)}{}$ in the case of $t_\mathrm{int}=\SI{4}{\micro\second}$, where the uncertainties are extracted errors from the fit.\\

\begin{figure}
    \centering
    \includegraphics{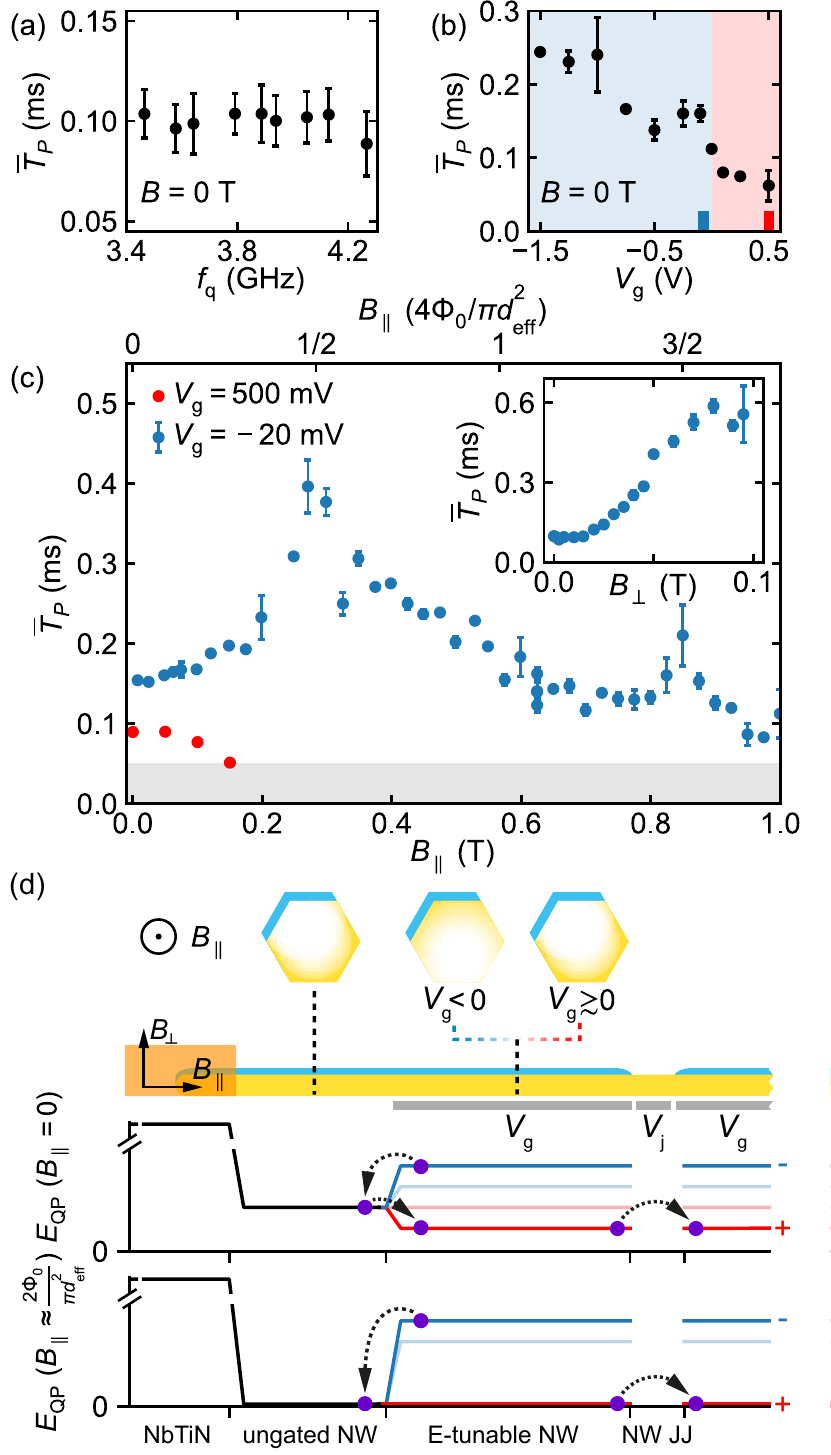}
    \caption{Parity lifetime dependence. (a) $\Tp$ versus $\fq$. (b) $\Tp$ versus $\Vg$. (c) $\Tp$ versus $\Bpar$ corresponding to the plunger settings marked by the ticks in b. $\Tp$ is not resolvable below $\Tp <\SI{50}{\micro\second}$ (grey area). Inset: $\Tp$ versus $B_{\perp}$. (d) Schematic illustrating the charge density profile in the cross-section of the NW, and its dependence on $\Vg$. The plunger gates locally alter the charge density distribution from ring-shaped for $\Vg\gtrsim0$ to charge confinement at the superconducting Al shell for $\Vg<0$. The minimal QP excitation energy $E_\mathrm{QP}$ is sketched for $B_\parallel=0$ (upper graph) and $B_\parallel\approx \frac{2\Phi_0}{\pi d_\mathrm{eff}^2}$ (bottom graph), the latter being the field required to thread half a flux quantum through the NW. At this field the orbital effect maximally suppresses the proximity-induced gap. Setting $\Vg< 0$ breaks the ring-shaped electron density profile, nullifying the orbital effect and restoring the proximity-induced gap in the gated sections. The field- and gate-dependent $E_\mathrm{QP}$ forms a potential landscape where QPs get trapped at specific locations in the wire, as illustrated.}
    \label{fig:Tp_dependencies}
\end{figure}
\noindent 
We now use the measurement of QPP dynamics to investigate the response of the device to static electromagnetic field tuning.
For these measurements ten individual measurements of $\Tqp$ are averaged over minute timescales to constitute $\Tp$.
The associated standard deviation in $\Tp$ captures any excess variation in $\Tqp$ stemming from temporal fluctuations in the QP density $\xqp$~\cite{Grunhaupt2018, Vool2014}.
First, we confirm that the observed QP dynamics result from the presence of non-equilibrium QPs, by measuring the temperature dependence of $\Tp$~\cite{SI}.
At the lowest temperature of $T\approx\SI{25}{\milli\kelvin}$ the extracted QP to Cooper pair ratio is $\xqp \approx 10^{-7}$, which corresponds to a thermal QP distribution with an effective temperature of $T_\mathrm{eff}^\mathrm{QP}\approx\SI{193}{\milli\kelvin}$. 
These numbers are typically encountered~\cite{Riste2013,Aumentado2004,Vool2014,Wang2014,Taupin2016}, and are linked to stray radiation impacting the device due to imperfect shielding~\cite{Serniak2019,Houzet2019,Vepsalainen2020}.

We take advantage of the semiconducting nature of the NW to electrostatically modify the charge-carrier density in the JJ and its leads~\cite{de_Moor_2018, Antipov2018, Winkler2019}.
First we verify that $\fq$ does not strongly influence the QP tunneling rate by varying $\fq$ through $\Vc$ [\Fref{fig:Tp_dependencies}(a)].
We measure a constant $\Tp$ while changing $\fq$ over almost \SI{1}{\giga\hertz}, corresponding to a \SI{3}{\milli\volt} increase in $\Vc$.
This justifies any retuning of $\fq$ amidst a measurement series, a necessary intervention to ensure that $\fq$ stays within a frequency range suitable for parity lifetime readout over large field ranges.
We then investigate the gate dependence of $\Tp$ by varying both plunger gate voltages $\Vg = \Visl= \Vgnd$ simultaneously [\Fref{fig:Tp_dependencies}(b)].
For $\Vg \gtrsim 0$ we observe a decrease in $\Tp$, while for $\Vg < 0$ we find $\Tp$ to rise quickly initially and more than doubles upon further decreasing $\Vg$.

The connection between $\xqp$ at the JJ and the proximity effect in the leads can be further explored by applying a magnetic field [\Fref{fig:Tp_dependencies}(c)].
For this we are able to operate the device as a charge parity detector up to $\Bpar=\SI{1}{\tesla}$, where $\Bpar$ is oriented along the NW.
For $\Vg \gtrsim 0$ we observe a decay of $\Tp$ with $\Bpar$~\cite{SI}, indicating increased $\xqp$ near the JJ as associated with the closing with magnetic field of the proximity-induced gap in the gated parts of the leads.
At $\Vg < 0$, we observe a markedly different dependence of $\Tp$.
Upon increasing $\Bpar$, $\Tp$ further rises and exhibits clear peaks at distinct values of $\Bpar$.
The transition point at $\Vg = \SI{0}{\volt}$ that divides these two regimes suggests that the device behaviour hinges on the energy potential landscape for QPs $\EQP$ set by the profile of the proximity-induced gap in the gated and ungated NW sections.

In the case of $\Vg < 0$, $\Tp$-enhancements indicate the turn-on of magnetic field induced QP trapping away from the JJ.
We observe enhancements of $\Tp$ for distinct values for $\Bpar$.
We attribute these to the orbital effect of the magnetic field, that occurs when the cylindrical electron density in the leads is threaded by magnetic flux, leading to the suppression of the proximity-induced gap [\Fref{fig:Tp_dependencies}(d)]~\cite{Buttiker1986,Kringhoj2021,Stampfer2021}.
Such a cylindrical electron density profile is a manifestation of charge accumulation at the InAs NW surface for $\Vg \geq 0$~\cite{Winkler2019}.
However, in the NW sections leading up to the JJ we can locally interrupt the accumulation by applying $\Vg < 0$, preventing such gap suppression.
The larger-gap sections obstruct QPs trapped in the ungated regions from reaching the JJ and thus acts as QP filters~\cite{Menard2019}.
The situation reverses upon setting $\Vg \gtrsim 0$, which activates the gap suppression and QP trapping right next to the JJ, decreasing $\Tp$ as $\xqp$ increases.
From the peak positions we estimate an effective diameter of $d_\mathrm{eff}=\sqrt{2\Phi_0/\pi \Bpar\left(\Phi_0/2\right)} \approx \SI{69}{\nano\meter}$ of the ungated NW sections, suggesting an accumulation layer of about \SI{13}{\nano\meter} based on the measured NW diameter of \SI{95(5)}{\nano\meter} and consistent with simulations~\cite{Antipov2018}.
Lastly, we apply an out of plane field $\Bperp$~[\Fref{fig:Tp_dependencies}(c) inset].
We now observe an even more dramatic dramatic increase of $\Tp$ at much more moderate fields, starting from $\Bperp > \SI{10}{\milli\tesla}$.
For perpendicular field orientation we do not expect the orbital effect to play a considerable role. Instead, we deem it more likely that the observed increase is caused by QP trapping in vortex cores induced in the NbTiN contacts~\cite{Wang2014,Taupin2016}, whose presence is also evidenced by the broadening of the resonator lineshape~\cite{Kroll2019}.
The resonator broadening prevented exploring higher ranges for $\Bperp> \SI{100}{\milli\tesla}$.

In conclusion, we have realized an OCS Al/InAs NW transmon that operates up to magnetic fields of \SI{1}{\tesla}, establishing a real-time charge-parity switching detector at strong magnetic fields.
We demonstrated a more than twofold increase in charge-parity lifetime by applying a local electrostatic potential and a global magnetic field, thereby controlling the minimal QP excitation energy in the leads of the JJ.
At negative gate voltages a QP filter forms at the JJ, trapping QPs away from the JJ.
The orbital effect suppresses the proximity-induced gap at half integer magnetic flux threading of the nanowire, increasing the efficiency of the QP traps and yielding magnetic field dependent parity lifetime enhancements.
At positive voltages and finite fields QPs are trapped at the JJ, resulting in significant drop in parity lifetime.
The demonstrated increase in parity lifetime via electric and magnetic field induced in-situ gap engineering, provides a realistic solution to QPP by non-equilibrium QPs, important for quantum computation with both topological and superconducting qubits.

\section*{Acknowledgements}
We thank K. Peterson, A. Kringh{\o}j, L. Casparis, T. Larsen, B. van Heck, D. Pikulin, T. Karzig, F. Rybakov, E. Babaev, K. Serniak, and B. Nijholt for the open and stimulating discussions, J. van Veen for depositing the NW and W. D. Oliver lab for providing a traveling wave parametric amplifier.

This work was supported by the Netherlands Organisation for Scientific Research (NWO/OCW) as part of the Frontiers of Nanoscience (NanoFront) program, Microsoft Corporation Station Q and the European Research Council.
\FloatBarrier
\bibliography{main}

\newpage
\onecolumngrid
\renewcommand{\thefigure}{S\arabic{figure}}
\setcounter{figure}{0}

\title{Supplemental materials for ``Field induced quasiparticle trapping in a hybrid superconducting-semiconducting circuit"}
\maketitle
\onecolumngrid
\newpage
\section*{Supplemental materials for ``Quasiparticle trapping by orbital effect in a hybrid superconducting-semiconducting circuit"}
\section{Device overview and experimental setup}
\FloatBarrier
As a substrate we use high resistivity silicon coated with \SI{100}{\nano\metre} of low pressure chemical vapor deposition silicon nitride (SiN$_x$) to prevent gate leakage at small gate separation. A thin \SI{26.5(1)}{\nano\metre} layer of NbTiN is sputtered onto the 3" wafer from a $99.99\%$ pure $\mathrm{Nb_{0.7}Ti_{0.3}}$ target in an \ch{Ar/N_2} atmosphere. We chose \ch{NbTiN} for its high upper critical field ($B_{c2}>\SI{8}{\tesla}$) and compatibility with microwave frequency circuits.
The coplanar waveguide (CPW) resonator, capacitor island, ground plane, bottom gate and LC-filters for the dc gate lines are all shaped from the base layer by reactive ion etching in a SF$_6$/O$_2$ plasma. The on-chip LC-filters are low pass filters that decouple the device from the off-chip circuitry at microwave frequencies.
In this same etching step the entire chip is patterned with artificial pinning sites to trap Abrikosov vortices~\cite{Kroll2019}.
The microwave resonator is asymmetrically coupled to its in- and output, where a lower coupling at the input prevents unnecessary losses.
To electrostatically isolate the bottom gate, \SI{32(1)}{\nano\metre} of SiN$_x$ is deposited by plasma-enhanced chemical vapor deposition and shaped by wet etching with buffered oxide etch (20:1). An indium arsenide (InAs) nanowire with epitaxial two-faceted thin-film aluminum (Al) shell~\cite{Krogstrup2015} is aligned with respect to the bottom gate using a micromanipulator setup.
The $\sim$\SI{100}{\nano\meter} long Josephson junction (JJ) is defined by a selective transene-D wet etch stripping the Al shell in the middle of the wire. Simultaneously a \SI{1}{\micro\metre} long part of the Al shell is also removed at the ends of the wire in preparation for the metallic contact patches.
\SI{177(5)}{\nano\metre} of \ch{NbTiN} is sputtered to contact both ends to the \ch{NbTiN} base, one end to ground and the other to the capacitor island. The \ch{NbTiN} patches overlap slightly with the remaining Al shell.
Local electrostatic tuning of the electron density in the wire is made possible by sputtering \SI{30(3)}{\nano\metre} of SiN$_x$ followed by $\sim$\SI{100}{\nano\metre} of sputtered NbTiN to form so-called wrap gates.

\begin{figure}[H]
    \centering
    \includegraphics[width=0.65\textwidth]{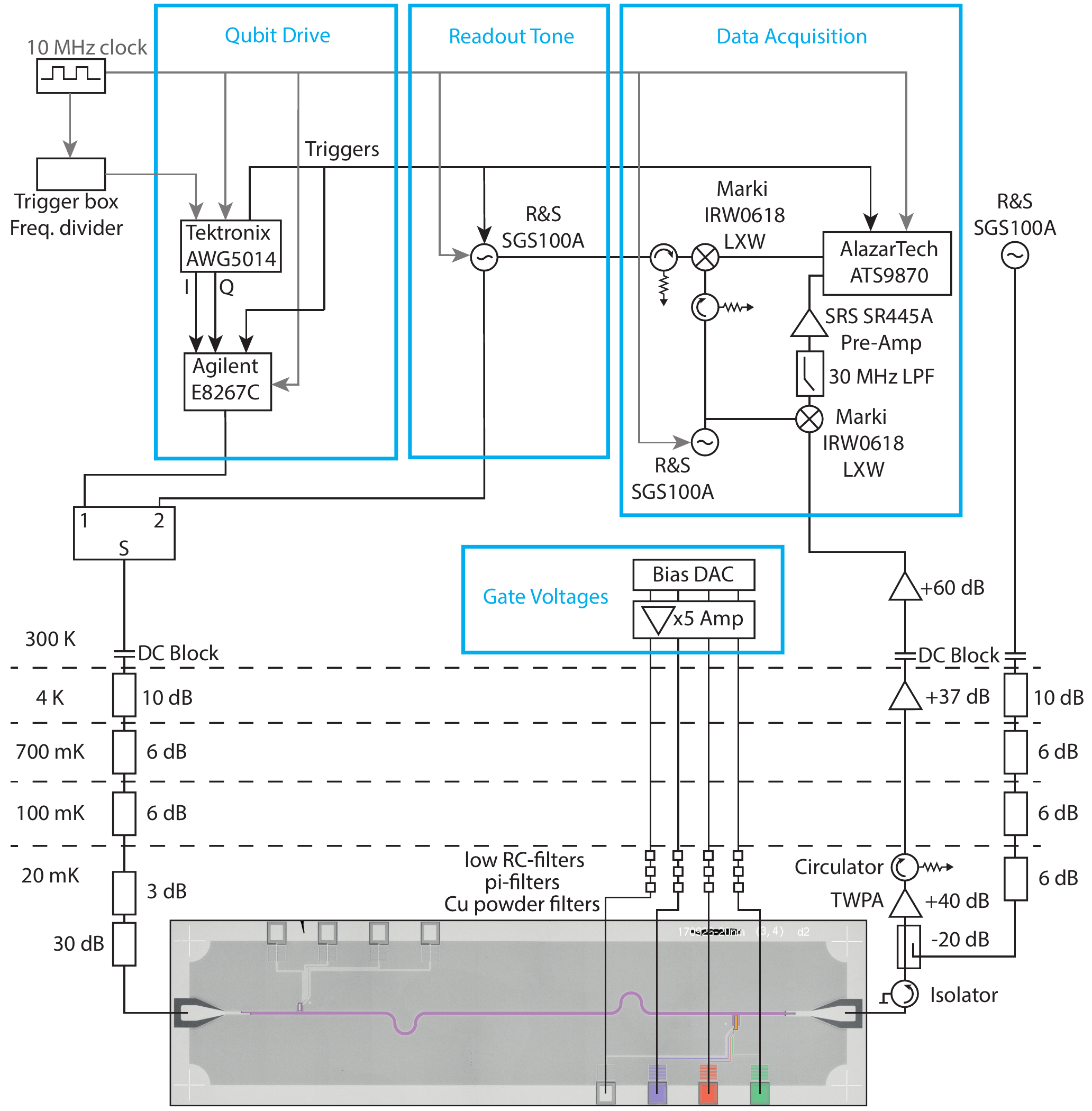}
    \caption{Schematic of the electronic measurement set-up including an optical image of the $\SI{2}{}\,\times\, \SI{7}{\milli\metre}$ sample. The overview of the sample shows the $\lambda/2$ CPW resonator and the LC-filter bondpad design.}
    \label{fig:setup}
\end{figure}
\FloatBarrier
\indent The chip is cooled down in a dilution refrigerator with base temperature of $\sim$\SI{25}{\milli\kelvin} and equipped with 3-axis vector magnet. 
To make place for the vector magnet the coldest radiative shield is mounted at \SI{700}{\milli\kelvin} and is not even light-tight.
To compensate for the lesser shielding the sample is placed in a copper casing encapsulated by several layers of eccosorb material and copper tape to absorb and reflect radiation.
Standard heterodyne measurement techniques are used to readout out the magnitude and phase response of the sample [\Fref{fig:setup}].
\indent To reduce the large parameter space the bottom gate was kept at \SI{0}{\volt} throughout these experiments.

\section{Transmon characterisation in Magnetic Field}
\begin{figure}[H]
    \centering
    \includegraphics[width=0.8\textwidth]{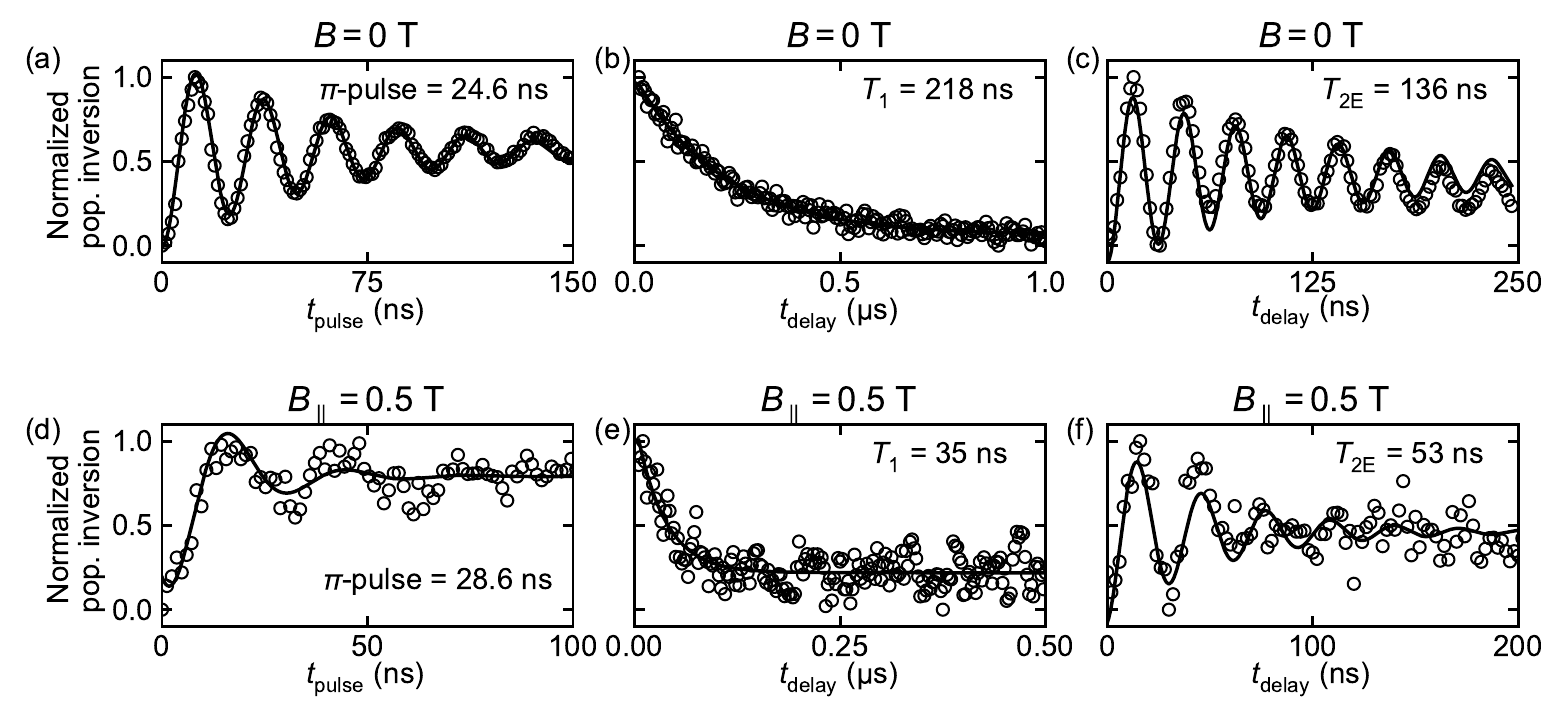}
    \caption{Time domain characterisation of the transmon. (a) Rabi oscillations at $B=$~\SI{0}{\tesla} showing coherent control of the 01-transition fitted by an exponential decreasing sinusoid (solid black). (b) Energy relaxation at $B=$ \SI{0}{\tesla} with exponential decaying fit corresponding to $T_1=$~\SI{218(6)}{\nano\second} (solid black). (c) Phase coherence at $B=$~\SI{0}{\tesla} measured by a Hahn-echo pulse sequence with an exponential decaying sinusoid corresponding to a time constant of $T_\mathrm{2E}=$~\SI{136(11)}{\nano\second}. (d) Rabi oscillations at $B=$~\SI{0.5}{\tesla} demonstrating magnetic field compatibility. (e) Energy relaxation at $\Bpar=$~\SI{0.5}{\tesla} with exponential fit (solid black) corresponding to a time constant of $T_1=$~\SI{35(3)}{\nano\second}. (f) Phase coherence at $B=$~\SI{0.5}{\tesla}, again measured by Hahn-echo pulse sequence. The decaying sinusoidal fit (solid black) yields $T_\mathrm{2E}=$~\SI{53(9)}{\nano\second}.}
    \label{fig:Suppl_TD}
\end{figure}

\begin{figure}[H]
    \centering
    \includegraphics[width=5in]{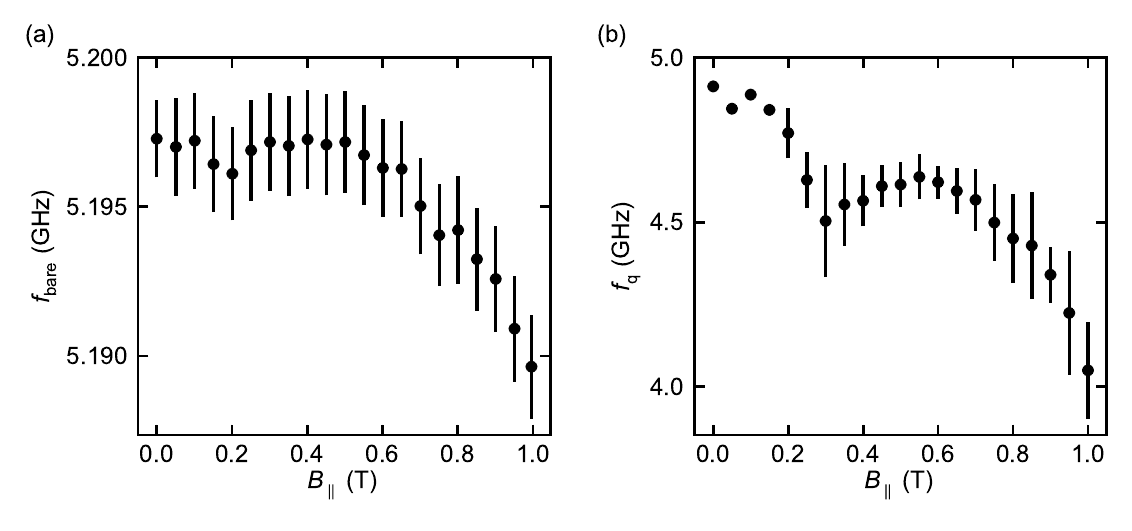}
    \caption{In-plane magnetic field compatibility of the circuit. (a) Resonance frequency of the decoupled resonator by reading out at high input power with the errorbars representing the full width at half maximum (FWHM). (b) Qubit frequency versus $\Bpar$, with the errorbars representing $4\times$FWHM for visibility. The FWHM ranges from \SIrange{25}{95}{\mega\hertz}.}
    \label{fig:Suppl_fq_vs_B}
\end{figure}

\FloatBarrier
To characterize the performance of the transmon we measure the energy relaxation and phase coherence times both at $\Bpar=$~\SI{0}{\tesla} and $\Bpar=$~\SI{0.5}{\tesla} [\Fref{fig:Suppl_TD}], demonstrating coherent control of the the transmon at large magnetic fields. The measurements at $B=$~\SI{0}{\tesla} correspond to a transmon frequency of $\fq=$~\SI{4.575}{\giga\hertz} with readout tone set to $\fp=$~\SI{5.206}{\giga\hertz} at gate settings $\Vc=$~\SI{255.9}{\milli\volt}, $\Visl=$~\SI{100.0}{\milli\volt} and $\Vgnd=$~\SI{100.0}{\milli\volt}. The parameter space where the transmon is measurably close to the resonator is significantly reduced at $\Bpar=$~\SI{0.5}{\tesla}. Therefore measurements were taken at different settings; $V_\mathrm{j}=$~\SI{260.0}{\milli\volt}, $\Visl=$~\SI{-418.6}{\milli\volt} and $\Vgnd=$~\SI{-419.0}{\milli\volt}, resulting in $\fq=$~\SI{4.580}{\giga\hertz} and $\fp=$~\SI{5.204}{\giga\hertz}. Both $T_{1}$ and $T_\mathrm{2E}$ times are significantly shorter than the observed parity lifetimes, showing that the coherence of our device is not limited by quasiparticle induced losses. Possible dominating decoherence mechanisms may be the lack of radiative shielding, unfiltered microwave lines and dielectric material losses.
\FloatBarrier
To further demonstrate the magnetic field compatibility we track the bare resonator response [\Fref{fig:Suppl_fq_vs_B}a] and the qubit frequency $\fq$ with $\Bpar$ [\Fref{fig:Suppl_fq_vs_B}b].
To ensure similar gate settings for the whole $\Bpar$-range, the measurement sequence was performed starting from $B_\parallel=\SI{1}{\tesla}$ at $\Vc=$~\SI{266.1}{\milli\volt}, $\Visl=$~\SI{-451.8}{\milli\volt} and $\Vgnd=$~\SI{-452.4}{\milli\volt}. 
Magnetic field alignment was performed at every magnetic field measurement to ensure the optimal alignment.
The bare resonator response, measured at high input power, shows a decreasing trend only from $\Bpar \sim$~\SI{0.4}{\tesla} on, leading us to believe this is the threshold for pinning all vortices in the resonator for  our design.
The slight dip in $f_\mathrm{bare}$ around $B_{ESR}\sim\SI{200}{\milli\tesla}$ is attributed to the dissipative loading of environmental paramagnetic electron spin resonances with Land\'{e} $g$-factor of $\sim2$~\cite{Kroll2019}. 
Two-tone spectroscopy of the simultaneously tracked qubit frequency reveals a non-monotonic decay.
The general decreasing trend is consistent with a decreasing supercurrent due the closing of the superconducting gap.
Comparison to [\Fref{fig:overview_Bparallel}] shows that $\fq$ depends more strongly on $\Bpar$ at more positive $\Visl, \Vgnd$ voltages.
This observation is in agreement with the density of states being confined towards the superconducting shell, improving the induced superconducting gap.
We observe a local minimum around $\Bpar = \SI{0.3}{\tesla}$, consistent with the flux induced interference effects as explained in the main text, and also reported by~\cite{Kringhoj2021,Stampfer2021}.
\FloatBarrier

\section{Validation of hidden Markov model}
\begin{figure}[H]
    \centering
    \includegraphics[width=0.8\textwidth]{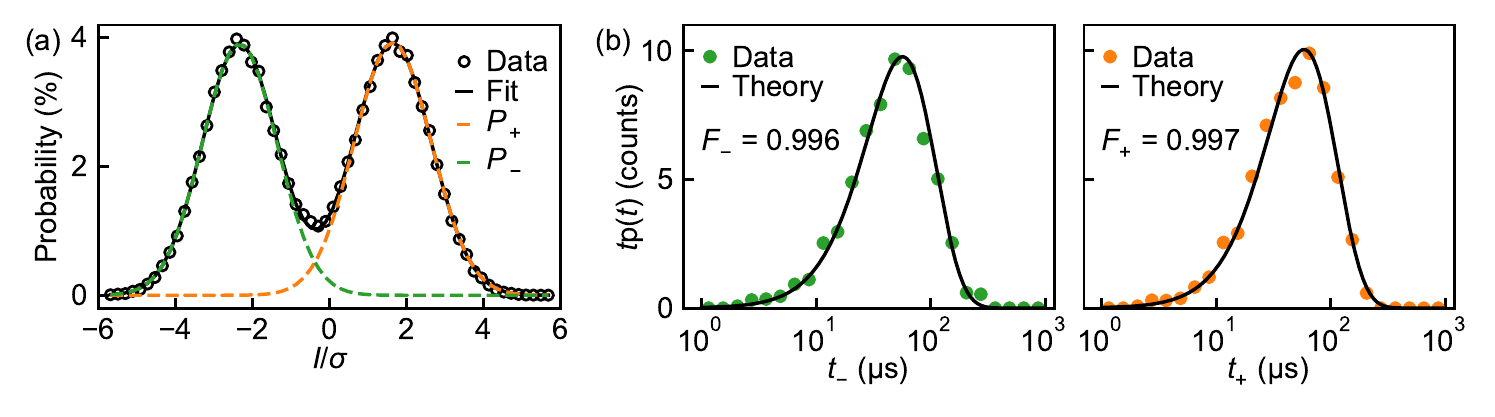}
    \caption{(a) Histogram projected on the I-quadrature (open black dots) fitted by a double Gaussian distribution (solid black) and decomposed into the two charge parity contributions (dashed green, dashed orange). (b) Histogram of dwell times for $P_{-}$ (left panel, green) and $P_{+}$ (right panel, orange) weighted by the dwell time, showing agreement with the theoretical predicted Poisson distribution (solid black). The fidelities $F_{-}$ and $F_{+}$ are extracted from the deviation of the data from the theoretical prediction.}
    \label{fig:Suppl_HMM}
\end{figure}

\FloatBarrier
As discussed in the main text, the assignment of a charge parity state to a data point is done by the use of a hidden Markov model (HMM)~\cite{Vool2014}.
To validate the use of a HMM to our data set, we show the two levels are normally distributed in the ($I,Q$) space [\Fref{fig:Suppl_HMM}(a)] and that the quantum jumps follow a Poisson distribution in time.
Deviation from a Gaussian distribution can indicate too long averaging per data point, not only averaging out noise but also averaging out the sharp jumps.
Secondly, the HMM is only applicable to Markovian processes that follow a Poissonian distribution in dwell time.
The exponential distribution shown in the main text verifies Poissonian behavior of the switching process. For a complete check, we verify the Poissonian behaviour also for the weighted histogram $tp(t)$ [\Fref{fig:Suppl_HMM}(b)]. Here the visibility of the low-frequency switches is increased. Comparison to the theory allows us to compute the fidelity:
\begin{equation}
    F = \frac{\sum_{i}\sqrt{M_{i}P_{i}}}{\sum_{i}M_{i}}
\end{equation}
where $M_{i}$ are the measured values and $P_{i}$ the predicted values. We find fidelities of $F_{-}=$~\SI{0.996}{} and $F_{+}=$~\SI{0.997}{}.
\FloatBarrier

\section{Parity lifetime extraction from PSD}
By use of the HMM we have shown symmetric transition dynamics, i.e. equal lifetimes and fidelities for the charge parity states.
This allows us to extract a combined charge parity lifetime $\Tqp$ with a combined detection fidelity $F$ by fitting the averaged power spectral density (PSD).
$F$ determines the detection error probability $(1-F)/2$, an uncorrelated noise process masking the random telegraph signal (RTS).
The transformation to the frequency domain is computationally beneficial, besides the SNR of the PSD can be increased while preserving high frequency information by averaging Fourier transforms of consecutive non-integrated time traces.
The PSD of a RTS is described by a Lorentzian~\cite{Riste2013}:
\begin{equation}
    S_{P} = F^2\frac{4\Gamma_{P}}{(2\Gamma_{P})^2+(2\pi f)^2} + (1-F^2)t_\mathrm{int}
\end{equation}
where $\Gamma_{P}=1/T_{P}$ is the average parity jump rate, $F$ the detection fidelity and $t_\mathrm{int}$ the integration time.
\FloatBarrier

\section{Effective quasiparticle temperature}
\begin{figure}[H]
    \centering
    \includegraphics[width=0.5\textwidth]{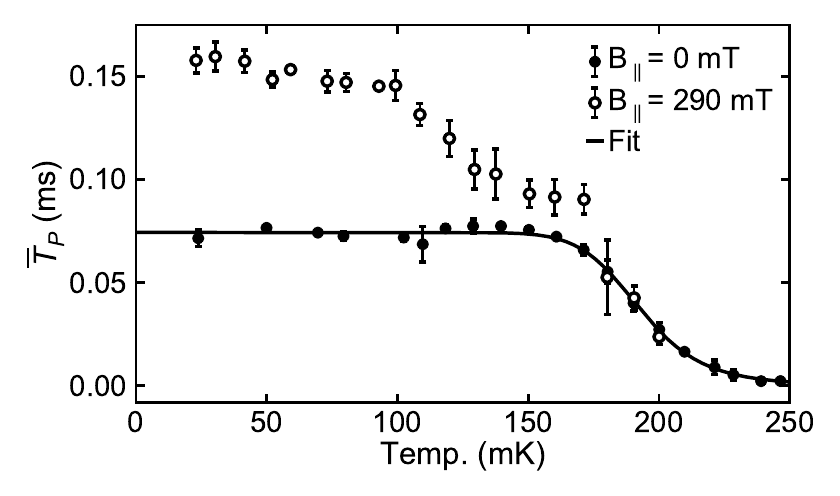}
    \caption{$\Tp$ versus temperature at $\Bpar =$~\SI{0}{\milli\tesla} (solid dots) and and $B_{\parallel} =$~\SI{290}{\milli\tesla} (open dots). The solid black line is a fit to the thermal dependence of $\Tp\propto\frac{1}{x_\mathrm{QP}}$ at $\Bpar =$~\SI{0}{\milli\tesla} which gives $x_\mathrm{QP}^{0} \approx 1 \times 10
^{-7}$, equivalent to an effective temperature of $T_\mathrm{eff}^\mathrm{QP} \approx$~\SI{193}{\milli\kelvin}, and $\Delta=\SI{257(16)}{\micro\electronvolt}$.}
    \label{fig:Temp}
\end{figure}

To investigate whether the main source of QPs are of non-thermal origin we measure the temperature dependence of $\Tp$, both at $\Bpar =$~\SI{0}{\milli\tesla} (at $\Vc=\SI{225}{\milli\volt}$, $\Visl=\SI{150}{\milli\volt}$, $\Vgnd=\SI{150}{\milli\volt}$) and at $\Bpar =$~\SI{290}{\milli\tesla} (at $\Vc=\SI{280}{\milli\volt}$, $\Visl=\SI{-100}{\milli\volt}$, $\Vgnd=\SI{-100}{\milli\volt}$) [\Fref{fig:Temp}].
$\Tp$ is determined from the Lorentzian fit of the frequency spectrum and we average 10 measurements to provide a standard deviation.
We assume that the QPP rate $\Gamma_P$ is proportional to the QP density $x_\mathrm{QP}$.
$x_\mathrm{QP}$ is taken as the QP density normalized by the density of Cooper pairs in thermal equilibrium and consists of two independent contributions:
\begin{equation}
    x_\mathrm{QP} = x_\mathrm{QP}^{0} + \sqrt{\frac{2\pi k_\mathrm{B}T}{\Delta}}e^{-\frac{\Delta}{k_\mathrm{B}T}}
\end{equation}
Here the first term covers the temperature independent contribution and the second term describes the temperature dependent contribution determined by scattering and recombination processes of QPs localized near the gap edge~\cite{Kaplan1976}.
The data is well-fitted by the model with $x_\mathrm{QP}^{0} \approx 1 \times 10
^{-7}$, equivalent to an effective temperature of $T_\mathrm{eff}^\mathrm{QP} \approx$~\SI{193}{\milli\kelvin}, and $\Delta =$~\SI{257(16)}{\micro\electronvolt}. Our result is similar to previously recorded QP ratios of $x_\mathrm{QP}^{0} \approx 10^{-8}$ to $10^{-5}$~\cite{Aumentado2004,Vool2014,Wang2014,Taupin2016}. Furthermore, $\Delta$ is consistent with DC measurements for thin aluminium films of \SIrange{7}{10}{\nano\metre}, indicating a near-perfect proximitized superconducting gap. 
To gain further insight at finite magnetic field we repeated the measurement at $\Bpar =$~\SI{290}{\milli\tesla} [\Fref{fig:Temp}].
A similar drop off around $\sim$\SI{190}{\milli\kelvin} shows the device is still plagued by the same source of non-equilibrium QPs. 
However, the cascading drop off can not be described by the theoretical model and therefore leads us to believe that the mechanism for $\Tp$-enhancement depends on temperature as well with a characteristic temperature that lies below \SI{190}{\milli\kelvin}.
\FloatBarrier

\section{Parity lifetime versus magnetic field}
\label{sec:TP_vs_B}
\begin{figure}[H]
    \centering
    \includegraphics[width=0.8\textwidth]{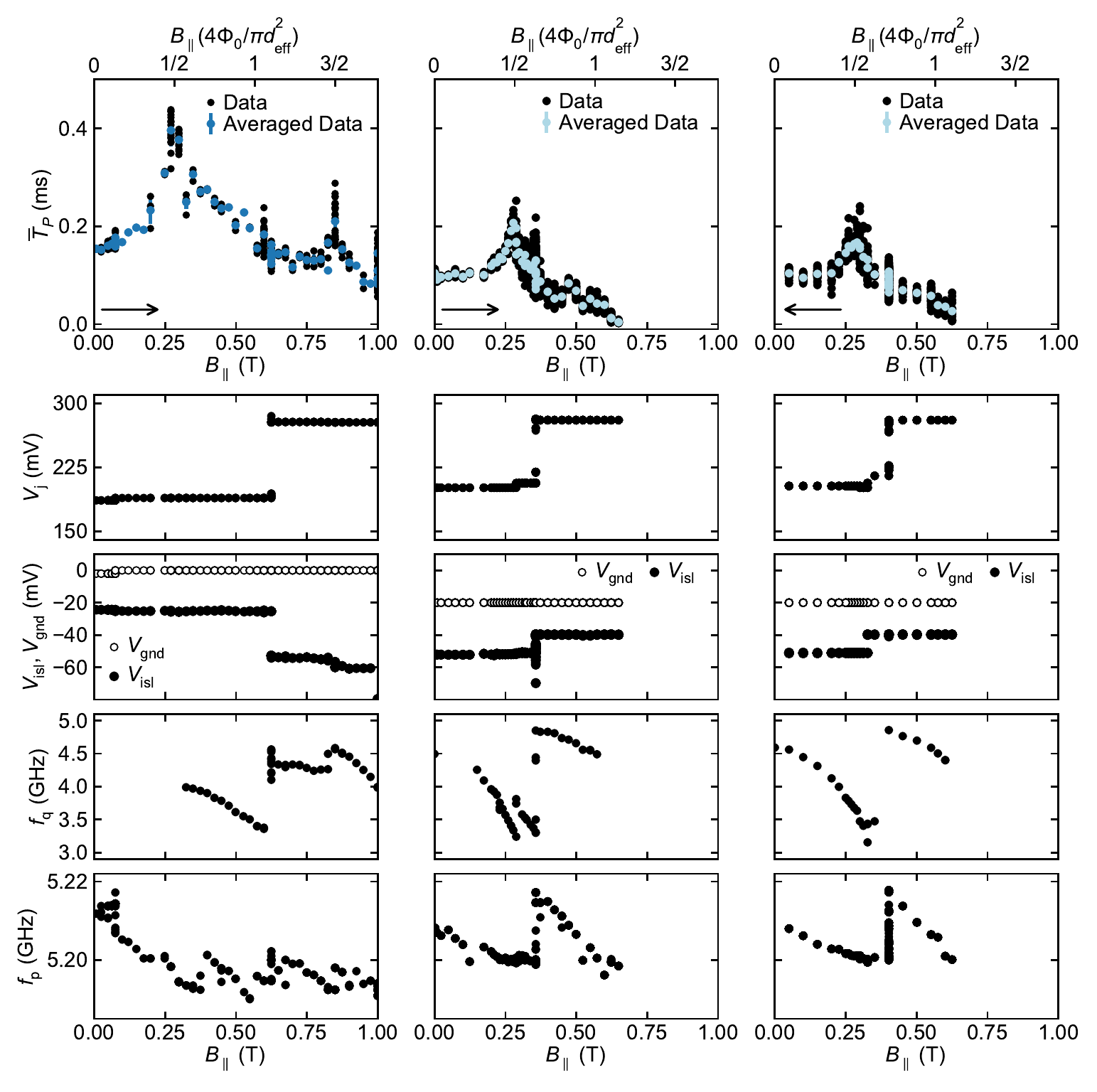}
    \caption{Overview of three measurement runs showing the in-plane magnetic field evolution of $\Tp$. The arrow in the bottom left corner indicates the sweep direction of the magnetic field. The corresponding gate settings $\Vc$, $\Visl$ and $\Vgnd$, readout frequency $\fp$ and qubit frequency $\fq$ are plotted below. The gate settings had to be retuned within a measurement run to stay within the detection range which is limited to $\fq=$\SIrange{3.2}{4.9}{\giga\hertz}.}
    \label{fig:overview_Bparallel}
\end{figure}

To demonstrate that the parity lifetime enhancement is a reproducible feature and is independent of the sweep direction, we repeated the measurement using different gate settings and sweep directions [\Fref{fig:overview_Bparallel}].
For completeness, all gate settings, readout frequency and available qubit frequency data are plotted along side.
The decreasing $\fq$ with $\Bpar$ shows the necessity of retuning some parameters during a measurement run, since the readout of $\Tqp$ is only possible in the range where $\fq=$\SIrange{3.2}{4.9}{\giga\hertz}.
We show here both the raw data and the averaged data, where averaging only has been applied to identical gate settings. We observe an enhancement in $\Tp$ for all three runs with the maximum at $\Bpar\simeq\SI{283}{\milli\tesla}$.

\begin{figure}[H]
    \centering
    \includegraphics{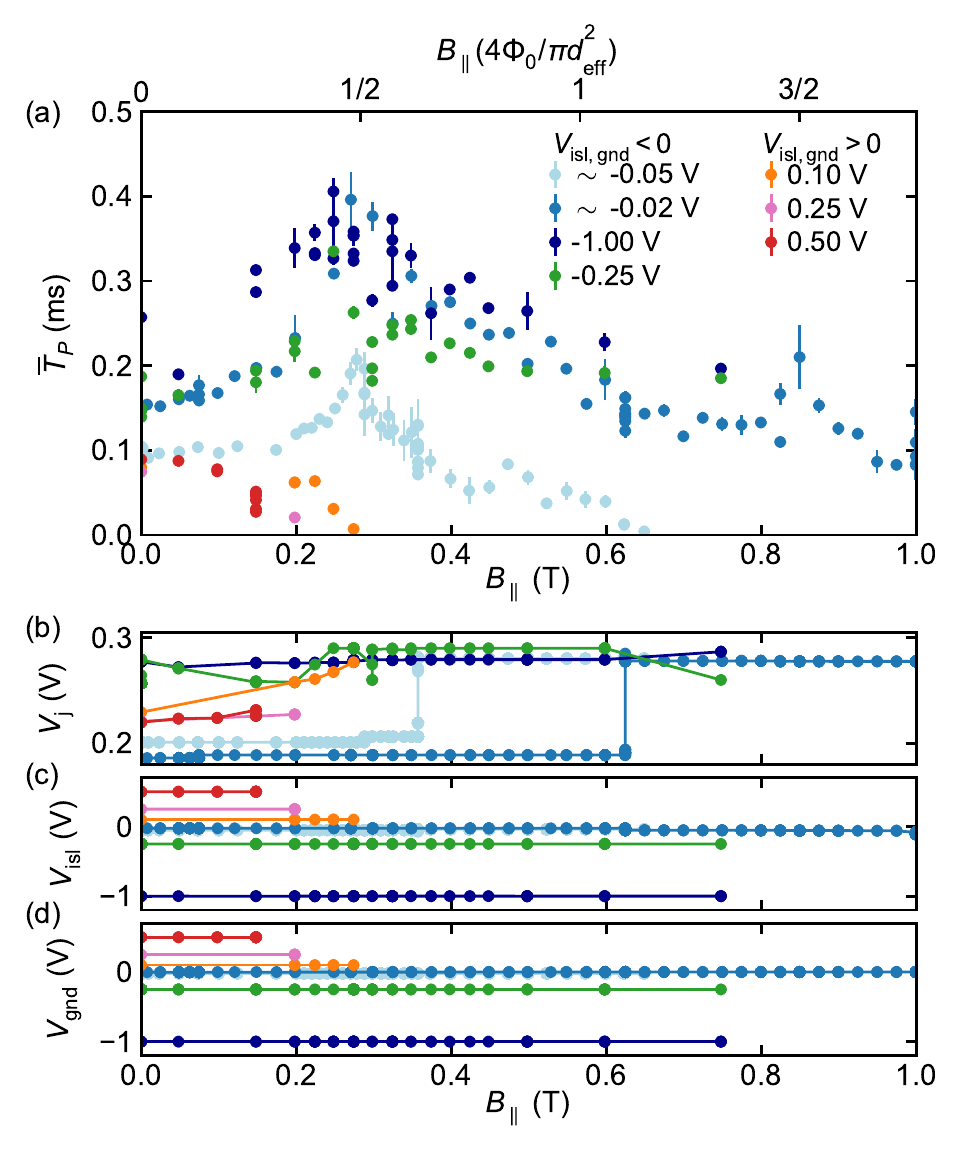}
    \caption{(a) $\Tp$ versus $\Bpar$ for different gate settings. Corresponding (b) $\Vc$, (c) $\Visl$ and (d) $\Vgnd$. Two markedly different trends have been measured, depending on the sign of the plunger gates $\Visl$ and $\Vgnd$. The top horizontal axis shows the magnetic flux penetration of the NW assuming $\Phi_0/2=\SI{283}{\milli\tesla}$.}
    \label{fig:test}
\end{figure}

Additional plunger settings are investigated to verify the difference between positive and negative $\Vg$ [\Fref{fig:test}].
The magnetic field alignment at each increment is time consuming. 
To save time, measurements of $\Tqp$ are taken by first sweeping over different gate settings before increasing $\Bpar$.
The measurement order results in a more capricious data set since the sweeping of gate voltages is prone to hysteresis.
Around \SI{280}{\milli\tesla}, we observe an enhancement of $\Tp$ for all $\Vg < \SI{0}{\volt}$, while we observe an impairment of $\Tp$ for $\Vg > \SI{0}{\volt}$.
The magnetic field value where the enhancement of $\Tp$ is observed seems identical.
Contrarily, the value at which $\Tp$ dips is observed to depend on the magnitude of the positive voltage applied to the plunger gates.
These observations suggest that the origin of the $\Tp$-enhancement is not located at the gated NW sections, while the phenomenon behind the $\Tp$-dips does occur at the gated NW sections.

\end{document}